\documentclass[12pt]{article}
\usepackage{graphicx}
\usepackage{amsmath}
\usepackage{mathabx}
\usepackage{authblk}

\newcommand{\degreesC}{$\;^\circ\text{C}\;$}

\title{The Resilience of Life to Astrophysical Events}

\author[1,a]{David Sloan}
\author[1,b]{Rafael {Alves Batista}}
\author[2,c]{Abraham Loeb}

\affil[1]{Department of Physics - Astrophysics, University of Oxford, Denys Wilkinson Building, Keble Road, OX1 3RH, Oxford, UK}
\affil[2]{Astronomy Department, Harvard University, 60 Garden Street, Cambridge, MA 02138, USA}

\affil[a]{david.sloan@physics.ox.ac.uk}
\affil[b]{rafael.alvesbatista@physics.ox.ac.uk}
\affil[c]{aloeb@cfa.harvard.edu}

\begin{document}

%\flushbottom
\maketitle

\begin{abstract}
Much attention has been given in the literature to the effects of astrophysical events on human and land-based life. However, little has been discussed on the resilience of life itself. Here we instead explore the statistics of events that completely sterilise an Earth-like planet with planet radii in the range $0.5-1.5 R_\Earth$ and temperatures of  $\sim 300 \; \text{K}$, eradicating all forms of life. We consider the relative likelihood of complete global sterilisation events from three astrophysical sources -- supernovae, gamma-ray bursts, large asteroid impacts, and passing-by stars. To assess such probabilities we consider what cataclysmic event could lead to the annihilation of not just human life, but also extremophiles, through the boiling of all water in Earth's oceans. Surprisingly we find that although human life is somewhat fragile to nearby events, the resilience of Ecdysozoa such as \emph{Milnesium tardigradum} renders global sterilisation an unlikely event.
\end{abstract}

\section{Introduction}

Within the coming years spectroscopy will likely be employed to identify molecules that are indicative of life in the atmosphere of exoplanets\cite{seager2016a}. In the context of the search for extraterrestrial life, it is useful to establish the necessary conditions for life to be present for such observations. Broadly speaking this relies upon two ingredients. The first is an unknown quantity -- the fraction of planets on which life begins. The causes of the emergence of life on Earth are not understood, and thus we do not have a complete theory for predicting where life may begin elsewhere. The second is the probability that life has persisted from its inception to observation. In this work we will show that this is highly likely, as events which could lead to life being completely eradicated are rare. To establish this we break from the usual study in the literature\cite{ruderman1974a,chapman1994a,dar1998a,beech2011a,piran2014a} of the possible paths to ending human life, and broaden the analysis to consider those astrophysical events which could rather remove all life by analysing the most resilient of species -- tardigrades. 

Tardigrades can survive for a few minutes at temperatures as low as -272\degreesC or as high as 150\degreesC, and -20\degreesC for decades\cite{hengherr2009a,tsujimoto2016a}. They withstand pressures from virtually 0 atm in space\cite{jonsson2008a} up to 1200 atm at the bottom of the Marianas Trench\cite{seki1998a}. They are also resistant to radiation levels $\sim 5000-6200 \; \text{Gy}$\cite{hashimoto2016a}. For complete sterilisation we must establish the necessary event to kill all such creatures.

We consider three types of astrophysical events which could constitute a threat to the continuation of our chosen life forms: large asteroid impact, supernovae (SNe), and gamma-ray bursts (GRBs). GRBs and SNe can be deadly due to the lethal doses of radiation and in particular the shock wave associated with the burst. Radiation can cause the depletion of the ozone layer, removing the shield that protects us from cosmic radiation\cite{ruderman1974a,ellis1995a,thorsett1995a}.

The effects of gamma-ray bursts (GRBs) on humans and land-based life could be disastrous as the eradication of the ozone layer would leave us exposed to deadly levels of radiation\cite{ruderman1974a}. However, in such circumstances life could continue below the ground. Significantly, several marine species would not be adversely affected, as the large body of water would provide shielding. Even the complete loss of the atmosphere would not have an effect on species living at the ocean's floor. The impact of a large asteroid could lead to an ``impact winter'', in which the surface of the planet receives less sunlight and temperatures drop. This would prove catastrophic for life dependent on sunlight, but around volcanic vents in the deep ocean life would be unaffected. Similarly, an increase in pressure, or acidity spread across the entirety of the (deep) ocean is an unlikely scenario for extinction. The physical processes by which ocean pressure could significantly increase involve increasing planetary mass; such impacts would first lead to extreme heating. Even following extreme events, spreading acidity through the entire ocean is unlikely. The removal of the atmosphere would also lead to mass extinction. However, following such an event the remaining ocean water would form a new atmosphere below which oceans could still form. The energy requirements for total sterilisation of the planet through atmospheric removal are significantly greater than those for boiling the oceans, so the threat of atmospheric removal is contained within that of oceans boiling. We are therefore led to consider death due to heat or radiation.

\section{Analysis}

To raise the entire ocean temperature by $T$ requires a deposit of $E=M_o \mathcal{C} T$ wherein $M_o$ is the ocean mass and $\mathcal{C}$ the specific heat capacity of water. In order to increase the temperature of the entirety of the Earth's oceans we need to introduce a large amount of thermal energy. The total mass of water in the oceans is around $1.35 \times 10^{21} \; \text{kg}$. 
The specific heat capacity of water is $4184 \; \text{J}\,\text{kg}^{-1}\,^\circ\text{C}^{-1}$
 so we require $5.6 \times 10^{24} \; \text{J}$ to raise the ocean temperature by 1\degreesC. Thus the tardigrade with a tolerance of up to 100\degreesC would survive until around $5.6\times 10^{26} \; \text{J}$ were deposited into the ocean. This is a lower bound -- such heat would not be evenly distributed, being it most likely to be deposited in the upper ocean. To provide a conservative bound, we seek to minimise the depth of the deepest ocean on any planet --  a uniform distribution of oceans across the planet's surface. When ocean mass is small compared to that of the planet, the depth of the ocean is approximately 
\begin{equation}
D=  \frac{\alpha \rho^{2/3} M_p^{1/3}}{(36\pi)^{1/3} \rho_w}.
\end{equation}
Here $\rho= 3M_p/(4\pi R_p^3)$ is the average planet density, $\alpha$ the fraction of the mass in ocean (on Earth $M_o \approx 2.3 \times 10^{-4} M_\Earth$) and $\rho_w$ the density of water. Most of Earth's water is contained within rocks. To remain conservative, we consider only the mass of liquid water in the oceans. There may exist planets that are almost entirely water ($\alpha \approx 1$), however for life as we know it, we focus on Earth-like planets with oceans on the surface of a rocky planet.
We give these explicitly as we will assume they are broadly unchanged between planets. For the Earth, this implies that there must be an ocean of at least 2.5 km in depth. This is far shallower than the deepest points, however it will constitute a lower bound. The intensity of gamma rays is attenuated by interaction with matter by a factor $\exp\left(-\mu D\right)$, wherein $D$ is the depth and $\mu$ the attenuation coefficient. This varies based on the material and the frequency of the incident radiation. The tardigrade is capable of withstanding over 6000 Gy (enough to endow every kilogram of material with 6000 J of energy). If the ocean depth is greater than $\frac{1}{\mu} \log(700)$ (the latter figure being the ratio of the energy deposit per unit mass required to boil water to that to kill a tardigrade) the water above will be boiling. In fact, if we consider a sufficient radiative flux to kill a tardigrade at depth $D$, the total energy deposited upon the planet is at least $E = \frac{6000 \pi R_p^2 }{\mu} (e^{\mu D} - 1)$. If our oceans are more than a few metres deep, this exceeds the threshold energy at which the oceans would boil before radiation would kill the tardigrade. We therefore consider temperature increase as the primary source of sterilisation. 

Large asteroids are the leading candidate for causing of the Cretaceous-Tertiary extinction which took place 65 million years ago, annihilating approximately 75\% of species on the planet leaving the Chicxulub crater. This event devastated larger land animals. Of those with masses over 25 kg only a few ectothermic species survived. However, around 90\% of bony fish species survived\cite{kriwet2004a} and deep ocean creatures were largely unaffected by the event. We estimate an upper bound for the energy deposited by an asteroid of mass $M_a$ as being its free-fall energy from infinity to the surface of the planet $E = 1 / 2 M_a (v_\infty^2 + v_e^2)$, where $v_e = \sqrt{2 G M_p / R_p}$ is the escape velocity of the planet ($v_e \approx 11.2 \; \text{km}\,\text{s}^{-1}$ for Earth), and $v_\infty$ is given by \"Opik's close encounter theory\cite{opik1976a}. In order to raise the ocean's temperature by $T$, we require an asteroid of mass
\begin{equation}
M_a = \frac{2 \alpha \mathcal{C} T}{v_\infty^2 + v_e^2} M_p.
\end{equation}
To annihilate tardigrades on Earth we require a mass over $\sim 1.7 \times 10^{18} \; \text{kg}$. The largest observed asteroids in the Solar System are Vesta and Pallas, with masses of $2.7 \times 10^{20} \; \text{kg}$ and $2.2 \times 10^{20} \; \text{kg}$ respectively. There are only 17 other known asteroids of sufficient mass, and a few dwarf planets, the most massive ones being Eris and Pluto, whose masses are $1.7 \times 10^{22} \; \text{kg}$ and $1.3 \times 10^{22} \; \text{kg}$ respectively. We reiterate that our estimate of the required energy is conservative -- it is likely that it would take a significantly more massive impact as ocean heat would only be a fraction of the total energy. Since we consider Earth-like planets, the order of magnitude of this mass does not vary greatly between the largest and smallest planets -- if oceans constitute an equal fraction of mass this changes by less than an order of magnitude. 

\begin{figure}
	\includegraphics[width=\columnwidth]{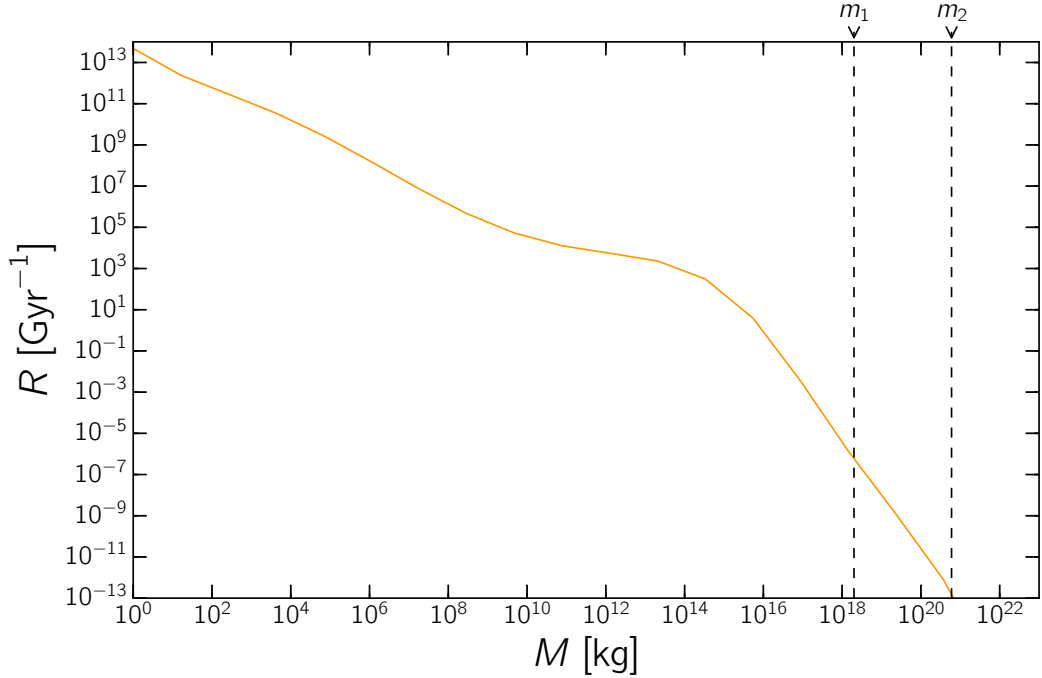}
	\caption{The cumulative impact rate per year for different masses of asteroids. This follows the parameterisation given in Ref.~\cite{bland2006a} up to $M = 10^{15} \; \text{kg}$; for $M > 10^{15} \; \text{kg}$ the impact rate is inferred following Ref.~\cite{hergarten2015a}. Dashed vertical lines indicate the minimum mass needed for complete sterilisation assuming a typical asteroid with density ($\rho = 2000 \; \text{kg/m}^3$). The lower bound ($m_1$) is that which could cause boiling of the oceans if the entirety of its energy were converted into heat spread homogeneously throughout the Earth's oceans. The upper bound ($m_2$) is the mass of an asteroid whose impact crater is equal to the size of the planet, causing complete destruction. Here we find that even with the most conservative bound, the likelihood of complete sterilisation is lower than around $10^{-5}$ over the lifetime of the planet.}
	\label{fig:asteroids}
\end{figure}

In figure~\ref{fig:asteroids} we present a model for the impact rate of asteroids as a function of the mass. This is based on the extrapolation of relation between crater diameter and impact rate\cite{hergarten2015a}. The mass of the object is related to crater diameter following Ref.\cite{collins2005a}, assuming asteroids with density $\rho \sim 5 \; \text{g}\,\text{cm}^{-3}$ entering the atmosphere with incidence angle of 90$^\circ$ with respect to the normal. In reality, most asteroids have $\rho \approx 2 \; \text{g}\,\text{cm}^{-3}$, and in the case of comets this value is even lower, being the value here assumed a conservative assumption. This is highly dependent on the asteroid distribution in our Solar System -- we assume in the absence of other evidence, that other systems are similar, however this remains to be verified.

The bulk of the energy output of a supernova is carried by the shock wave. To give an upper bound on the range at which a supernova would remove all life from the planet, we assume that the shock wave carries all the energy released. The fraction of energy incident on a planet of radius $R_p$ at a distance $d$ from the supernova is given by the fraction of the sphere of radius $d$ which is covered by the planet's surface $\frac{\pi R_p^2}{4 \pi d^2}$. To raise the temperature of a planet by $T$, we would require a supernova within a distance $d$ given:
\begin{equation}
d_{SN} = \left(\frac{3}{32 \pi M_p^{1/2} \rho} \right)^{\frac{1}{3}} \left( \dfrac{E_s}{\alpha \mathcal{C} T} \right)^\frac{1}{2}
\end{equation}
For the Earth, this sterilisation distance is around 0.04 pc, far closer than the closest stars, Proxima Centauri. Were a supernova to occur at that distance, the ocean temperature would only rise by about 0.1\degreesC. Furthermore, although there is a dependence on the mass of the planet, this dependence is quite weak. Note that none of the stars in the Alpha Centauri system are large enough to go supernova. The nearest potential supernova is the IK Pegasi system, approximately 45 pc away, which is three orders of magnitude farther than the estimated sterilisation radius.

\begin{figure}
	\includegraphics[width=\columnwidth]{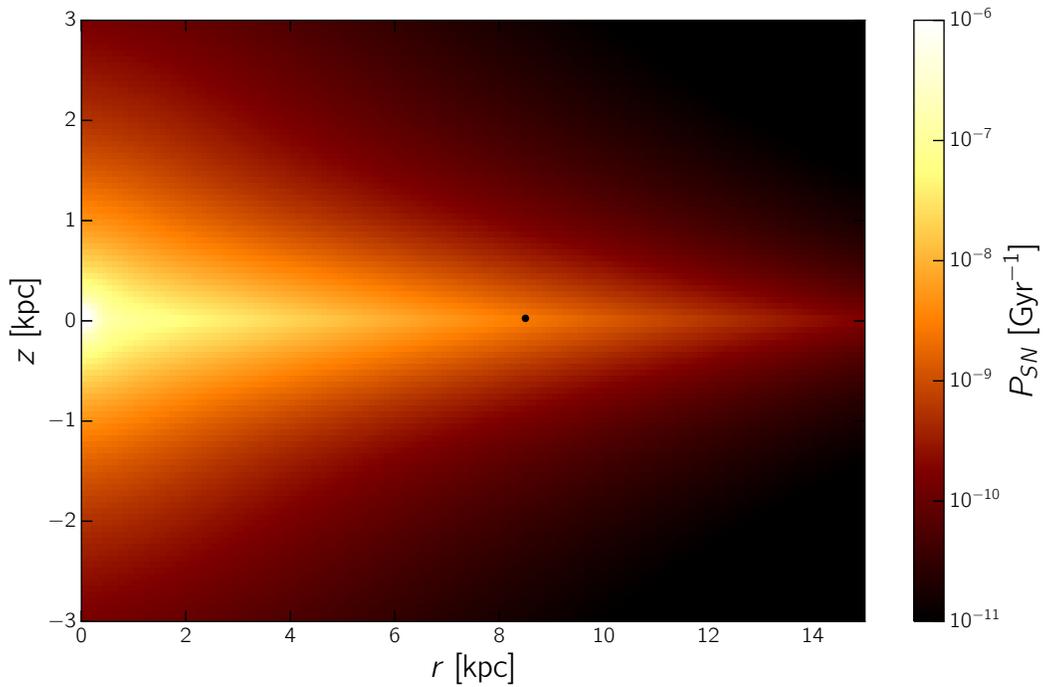}
	\caption{The expected number of supernovae within the sterilisation sphere of radius 0.04 pc per Gyr as a function of galactic position. The black circle indicates the position of the Solar System. Closer to the galactic centre, the stellar density is higher, and thus the likelihood of encountering a nearby supernova increases. However, this density is only sufficient to give a total rate of around 0.01 expected events per billion years, and thus total sterilisation through supernovae is still an improbable event.}
	\label{fig:SNrates}
\end{figure}

To assess the relative risk faced by any planet in our galaxy, we approximate the odds of a close enough supernova happening over a timespan of $10^9$ years. We find the expected number of stars of sufficient mass within the sterilisation distance of a planet, and the odds that one of these stars goes supernova. The galactic habitable zones, regions wherein complex life may evolve, depends on the occurrence rate of supernovae. A detailed simulation-based study was done by Lineweaver {\it et al.}\cite{lineweaver2004a}. We evaluate the rate of SN at a position $(r, z)$ (cylindrical coordinates) as follows:
\begin{equation}
	P_{SN}(r,z) = \chi \int_{M_{min}}^{M_{max}} dm \, \xi(m) n_\star(r,z) \tau^{-1}(m), 
\end{equation}
with $n_\star$ being the number density of stars~\cite{juric2008a}. Following Ref.\cite{juric2008a} we select $M_{min}=8M_\odot$ and $M_{max}=25M_\odot$ for supernova progenitors. $\xi(m)$ and $\tau^{-1}(m)$ are respectively the initial mass function and the lifetime of a star of mass $m$. This is normalised to the global supernova rate in the Milky Way\cite{tammann1994a}. The rate of supernovae explosions within the sterilisation radius (0.04 pc) over 1 billion years is shown in figure~\ref{fig:SNrates}, for differing galactic locations.

Because the nearest star is about 1.3 pc away, we can conclude that Earth is located in a fortunate position. Near the galactic centre the density of stars increases and the probability of a SN sterilising life is higher. Nonetheless, this rate is almost insignificant even close to the galactic core, reaching only around 1\% of planets being sterilised. % check

The calculation for GRBs is similar to the one for SNe, but now we have to assume that the energy is collimated into in jets. As before, we assume the most pessimistic scenario -- the smallest jet angle with the largest energy. The energy is typically the same as that of a supernova, $10^{44} \; \text{J}$, but the jet angles can be as low as $2^\circ$, hence the energy incident on a target of radius $r$ which lies entirely within the beam angle at a distance $d$ is:
\begin{equation}
E = E_{GRB} \frac{\pi r^2}{\Omega d^2} = \frac{10^{62 } r^2}{d^2} \;\text{J}
\end{equation}
Hence for an increase of 100\degreesC in the ocean temperatures, we would need a GRB within about 13.8 pc; again, this is an upper limit.
% -- for larger beam angles, only a fraction of the energy being deposited in the ocean etc would mean a nearer object was required. 
The rate of occurrence of short GRBs per volume in the universe is $0.04 \; \text{Gpc}^{-3} \; \text{yr}^{-1}$, and long GRBs is $0.15 \; \text{Gpc}^{-3} \; \text{yr}^{-1}$\cite{piran2014a}. We will restrict these occurrences to within galactic discs of stars, therefore we divide this by the product of the comoving number density of galaxies  ($\approx 10^{7} \; \text{Gpc}^{-3}$) and the volume occupied of the galactic disc ($10^{11} \: \text{pc}^3 $), we find that the rate is around $2\times 10^{-10}  \; \text{pc}^{-3} \; \text{Gyr}^{-1}$, and hence the probability of a GRB within the a distance at which it would sterilise a planet, aligned such that one of the beams hit the planet is $3.2 \times 10^{-10} \; \text{Gyr}^{-1}$. This number is extremely small and we can conclude that such event is unlikely.

Planetary systems can also be disrupted by passing-by stars. The rate, $R$, of such encounters for a given stellar density, $n_\star$, is
\begin{equation}
	R = n_\star \sigma v,
\end{equation}
where $\sigma$ is the cross section to disrupt the orbit of the planet, and $v$ the velocity of the star. Typically, $v \approx 40 \; \text{km s}^{-1}$. In the neighbourhood of the Solar System $n_\star \sim 10^{-3} \; \text{pc}^{-3}$~\cite{juric2008a}. In the case of Earth~\cite{laughlin2000a}, $\sigma \sim 10^{-9} \; \text{pc}^{2}$. Therefore, the rate of interactions of stars and the Earth-Sun system would be $R \sim 3\times 10^{-8} \; \text{Gyr}^{-1}$. If we repeat this calculation for the average stellar density in the galaxy, which is $\sim 0.1 \; \text{pc}^{-3}$, the rate would be increased to $\sim 10^{-6} \; \text{Gyr}^{-1}$. This number is an upper bound -- we expect that only a fraction of systems that experience disruption would eject a planet -- yet it is still extremely small and we can conclude that ejection by this mechanism is a very rare event.  

\section{Discussion}

Our analysis has focussed on providing an absolute upper bound for the rate of complete sterilisation of an Earth-like planet during its evolution, by considering the required events that would lead to the death of the hardiest species on Earth. With such assumptions, we find that the probability is less than $10^{-7}$ per billion years. The overall likelihood of complete sterilisation is small even for planets which could exist around dwarf stars for ten trillion years, the most likely time for life to find itself\cite{loeb2016a}. For asteroids the impact rate of deadly objects is $\lesssim 10^{-5} \; \text{Gyr}^{-1}$. Eventually the evolution of the host star will render a planet sterile, either through expansion to the point of oceans boiling, or through a post-collapse freezing. There is a third scenario, where life continues around geothermal vents on a rogue planet until capture by a new host system, or the source of heat is extinguished. The time scale for the former is conservatively bounded by the rate of disruption of planetary systems by wandering stars, being therefore $\sim 30 \; \text{Myr}$ in regions with density of stars comparable to the vicinity of the Solar System; the time scale for the latter depends on properties of the planet, but are of the order of billion years. Consequently, life could perdure on a rogue planet long enough for it to be recaptured. We do not fully understand the mechanisms by which life started, but once it exists on an Earth-like planet, the complete removal of all life (other than through evolution of the host star) is a very unlikely event. 

In this analysis we have considered only direct effects of astrophysical events such as the direct boiling of oceans; however such events could serve as a trigger for a second change. An example of this is the runaway greenhouse effect~\cite{goldblatt2015a} in which a less energetic event could evaporate a fraction of the ocean, which in turn leads to increased temperatures. Such an occurrence would have a lower threshold energy, as total evaporation of the oceans is no longer necessary; however the precise details of such an event are not known. As an upper bound, we can presume that this would have to increase global temperatures by a larger amount than seasonal variations and changes in solar output. In essence this would reduce our energy threshold by about two orders of magnitude, still leaving such an occurrence very rare. 
% The strong greenhouse effect on, e.g. Venus, is much more likely to be the result of local environmental changes than an astrophysical event.}

Throughout this analysis we have made several assumptions, the primary being that life elsewhere will be similar to that found on Earth. To justify this assumption, we consider that when searches are conducted for exoplanetary life, the template used is that of life on Earth. We make the further assumption that life will evolve to adapt to the extreme environments of exoplanets as it has to those on Earth. Again, we justify this by the ubiquity of life across environmental conditions. 

Finally, we note that the type of life we expect to survive all but the most extreme of events is that which could survive elsewhere in our own solar system. The history of Mars indicates that it had an atmosphere\cite{jakosky2015a} that could have supported life, albeit in conditions that on Earth would be considered inhospitable, with both the \emph{Mars 2020} and \emph{ExoMars} programs aiming for experimental verification. Organisms with similar tolerances to radiation and temperature such as tardigrades would be the only kind of life that could survive long-terms in such conditions, and even then they would have to be significantly below the surface. The subsurface oceans that are posited to exist on Europa and Enceladus would have conditions similar to the deep oceans of Earth where taridgrades are found - volcanic vents providing heat in an environment devoid of light. The presence of extremophiles in the locations would be a significant step in narrowing the question of life starting on exoplanets. 

% \begin{methods}

% We estimate an upper bound for the energy deposited by an asteroid of mass $M_a$ as being its free-fall energy from infinity to the surface of the planet $E = 1 / 2 M_a (v_\infty^2 + v_e^2)$, where $v_e = \sqrt{2 G M_p / R_p}$ is the escape velocity of the planet ($v_e \approx 11.2 \; \text{km}\,\text{s}^{-1}$ for Earth), and $v_\infty$ is given by\cite{opik1976a}:
% \begin{equation}
% 	v_\infty^2 = \dfrac{G M_s}{a_\Earth} \left[  3 - \dfrac{a_\Earth}{a} - 2 \sqrt{\dfrac{a}{a_\Earth}(1 - e^2) } \cos \theta_i  \right],
% \end{equation}
% where $a$ is the semi-major axis, $e$ the orbit eccentricity, $\theta_i$ the inclination of the orbit, and $M_s$ the mass of the central star ($M_s = M_\odot$ for the Solar System). If we let $\alpha$ be the fraction of the mass of the planet that is oceans (on Earth $M_o \approx 2.3 \times 10^{-4} M_\Earth$), then we find that in order to raise the ocean's temperature by $T$, we require an asteroid of mass
% \begin{equation}
% M_a = \frac{2 \alpha \mathcal{C} T}{v_\infty^2 + v_e^2} M_p.
% \end{equation}

% The impact rate for craters with a given diameter is given by\cite{hergarten2015a}. Following 

% % \subsection{Method subsection.}
% % Here is a description of a specific method used.  Note that the subsection heading ends with a full stop (period) and that the command is \verb|\subsection{}| not \verb|\subsection*{}|.

% \end{methods}

\section*{Acknowledgements}

D.S. and R.A.B. acknowledge the financial support from the John Templeton Foundation. 

\section*{Author contributions statement}

\noindent The idea of this work was conceived by A.L. 
\noindent D.S. and R.A.B. contributed equally to the analysis of results, with some input from A.L. D.S. did most of the writing, with aid of R.A.B. Figures were produced by R.A.B. 
\noindent A.L. was responsible for the scope and accuracy checking of the analysis.

\section*{Additional information}

The authors declare no competing financial interest.

\bibliographystyle{unsrt}
\bibliography{references}

\end{document}